\documentclass[12pt,preprint]{aastex}

\shortauthors{Kilic et al.}
\shorttitle{White Dwarf GD362}
\begin{document}
%\title{A Parallax Distance for the Possibly Massive White Dwarf GD362\altaffilmark{1}}  
\title{Direct Distance Measurement to the Dusty White Dwarf GD 362\altaffilmark{1}}

\author{Mukremin Kilic\altaffilmark{2,3}, John R. Thorstensen\altaffilmark{4}, and D. Koester\altaffilmark{5}}

\altaffiltext{1}{Based on observations obtained at the Michigan-Dartmouth-MIT (MDM) Observatory, operated by
Dartmouth College, Ohio State University, Columbia University,
the University of Michigan, and Ohio University.}

\altaffiltext{2}{Spitzer Fellow, Smithsonian Astrophysical Observatory, 60 Garden Street, Cambridge, MA 02138;\\ mkilic@cfa.harvard.edu}

\altaffiltext{3}{Department of Astronomy, The Ohio State University, 140 West 18th Avenue, Columbus, OH 43210}

\altaffiltext{4}{Department of Physics and Astronomy, Dartmouth College, 6127 Wilder Laboratory, Hanover,
NH 03755; john.thorstensen@dartmouth.edu}

\altaffiltext{5}{Institut f\"ur Theoretische Physik und Astrophysik, University of Kiel, 24098 Kiel, Germany}

\begin{abstract}

We present trigonometric parallax observations of GD 362 obtained over seven epochs using
the MDM 2.4m Hiltner Telescope. The existence of a dust disk around this possibly massive white
dwarf makes it an interesting target for parallax observations. The measured parallax for GD 362
places it at a distance of $50.6^{+3.5}_{-3.1}$ pc, which implies that its radius and mass are
$\approx 0.0106R_\odot$ and $0.71M_\odot$, respectively.
GD 362 is not as massive as initially thought (1.2$M_\odot$).
Our results are entirely consistent with the distance and
mass estimates (52.2 pc and 0.73$M_\odot$)
by Zuckerman et al., who demonstrated that GD 362 has a helium
dominated atmosphere. Dropping GD 362 from the list of massive white dwarfs, there are no white dwarfs with
$M>0.9M_{\odot}$ that are known to host circumstellar dust disks.  

\end{abstract}


\section{Introduction}

\citet{gianninas04} reported the discovery of the most massive (1.2$M_{\odot}$), hydrogen- and metal-rich DAZ white dwarf
ever found; GD362 (WD 1729+371). In addition to Balmer lines, they detected \ion{Ca}{1}, \ion{Ca}{2}, \ion{Mg}{1}, and \ion{Fe}{1} lines in
the optical spectra of this star with estimated $T_{\rm eff}=$ 9740 K and $\log$ g = 9.1. They measured nearly-solar abundances
for these metals. The discovery of infrared excess \citep{kilic05,becklin05} and a silicate emission feature at
$\approx 10~\mu$m \citep{jura07} demonstrated that GD 362 hosts a circumstellar dust disk and the metals are likely to be
accreted from this disk.

The discovery of debris disks around other normal mass white dwarfs \citep[$M<0.8M_\odot$,][]{kilic06,kilic07,vonhippel07,
jura07,farihi08a}
and the lack of disks around massive white dwarfs with $M>1M_\odot$
\citep{hansen06} showed that, if GD 362 is as massive as 1.2$M_{\odot}$, it is unique. \citet{garcia07} proposed
to explain the high mass of GD 362 by a merger of two lower mass white dwarfs, a rare event in our Galaxy
\citep[see also][]{livio92,livio05}.
They also suggested that the merger scenario is able to explain the observed photospheric composition of metals and
the infrared excess of the surrounding debris disk.

A caveat in the model atmosphere analysis, and therefore surface gravity and mass determinations, of GD 362 is the presence of helium.
Helium lines become invisible below $\approx$ 11,000 K, and weak lines can only be seen if helium is present in significant amounts.
\citet{garcia07} found that the Balmer line profiles in the optical spectrum of GD 362 are equally well reproduced
even if there is a significant amount of helium in the atmosphere. No helium lines were visible in the low resolution spectrum
of \citet{gianninas04}. However, a high resolution and high signal-to-noise spectrum of GD 362 obtained by \citet{zuckerman07}
revealed a helium absorption line at 5876 \AA, demonstrating that helium is present in significant amounts.
A detailed model atmosphere analysis of this spectrum implied that GD 362 has $T_{\rm eff}=$ 10,540 $\pm$ 200 K, $\log$ g = 8.24
$\pm$ 0.04, $\log$ [He/H] = 1.14 $\pm$ 0.10, and $M=0.73M_\odot$; GD 362 may not be unique (in terms of its mass) after all. 

The high mass measurement for GD 362 implies a distance of 22-26 pc \citep{gianninas04}, whereas the lower mass solution found
by \citet{zuckerman07} implies twice the distance. Discriminating between these two solutions requires parallax observations.
In this Letter, we present trigonometric parallax observations of GD 362. Our observations and reduction procedures are discussed
in \S 2, while the observed parallax and its implications are discussed in \S 3 and \S 4.

\section{Parallax Observations and Reductions}

Our parallax images are from the 2.4m Hiltner reflector at MDM Observatory
on Kitt Peak, Arizona.  We used a total of 117 images taken on seven observing 
runs spanning two seasons; Table 1 summarizes the observations.  
The instrumentation, observing protocols, and reduction
techniques used were very similar to those described in \citet{thor03} and
\citet{thor08}.  We used a 2048$^2$ SITe CCD detector at the f7.5 focus;
each 24 $\mu$m pixel subtended $0''.275$.  At each epoch we took many exposures
in the Kron-Cousins $I$-band, as near to the meridian as we could to minimize
differential color refraction (DCR) effects \citep{monet92}.  There were
some differences from the earlier work, namely (1) we used a 4-inch 
filter, which allowed us to use the full imaging area of the detector; and 
(2) the chip was oriented with the columns east-west rather than north-south.
The parallax reduction and analysis pipeline was unchanged from the previous work.
In the GD362 field, we measured 83 stars, and used 43 of them to define the 
reference frame. 

Standardized magnitudes and colors are used in the parallax analysis, 
to correct for DCR effects and also to estimate the correction from relative 
to absolute parallax \citep{thor03}.  We took images of GD362 in $V$ (as well 
as the $I$ used for astrometry) on two observing runs (2008 June and September), 
and calibrated the instrumental magnitudes from these images with numerous 
observations of \citet{landolt92} standard star fields 
from the same runs.  The photometry of the GD362 field from the 
two runs agreed within a few hundredths of a magnitude in $V$ and $V-I$.  

To arrive at a final distance estimate we use procedures detailed in 
\citet{thor03}. For GD362, we measure a proper motion of $\mu=211.8 \pm 2.0$ mas (milli-arcsec) yr$^{-1}$
with an angle $\theta= 173.5^o \pm 0.4^o$,
and we find a relative parallax and formal error of $19.0 \pm 0.7$ mas.
Our proper motion measurement is relative to the chosen reference
stars, and is not reduced to an absolute reference frame. Corrections to absolute proper
motions are generally of order $<$ 10 mas yr$^{-1}$ \citep{lepine05}, and our measurement is
consistent with $\mu=224.2 \pm 7.8$ mas yr$^{-1}$ found by \citet{salim03}.
Figure 1 displays the trajectory of GD 362 on the sky, and the same trajectory
with proper motion component taken out. This figure shows that our observations cover
a large range of parallax factor for GD 362, and the parallax is well constrained.
Using the scatter observed among the parallaxes of comparably bright field
stars located near the program star we estimate an external error of 1.3 mas. 
The magnitudes and colors of the reference stars yield a correction 
to absolute parallax of 0.9 mas.  Combining these gives an
absolute parallax and external uncertainty  $\pi_{\rm abs} =
19.9 \pm 1.3$ mas, corresponding to 
$50.3^{+3.5}_{-3.1}$ pc.  The parallax accuracy is good enough that 
further corrections have only a small effect on the distance.  
Nonetheless, the full Bayesian formalism described in \citet{thor03} 
adjusts this slightly, to $50.6^{+3.5}_{-3.1}$ pc.  This formalism
includes (1) the Lutz-Kelker correction \citep{lutzkelker}, which
increase the inferred distance to 51.0 pc, and (2) prior information
from the proper motion and very liberal limits on the luminosity,
which work to decrease the distance slightly.

\section{Results}

Our parallax measurement shows that GD 362 is further away, and therefore more luminous and less massive, than predicted
from the pure hydrogen atmosphere fits by \citet{gianninas04}. We now compare our result to the distance implied by
the model atmosphere fits by \citet{zuckerman07}.

GD 362 is included in the Sloan Digital Sky Survey (SDSS) Data Release 6 imaging area. We converted the SDSS photometry
into the AB system using the offsets given in \citet{eisenstein06}. GD 362 has $u=16.25,~g=16.02,~r=16.18,~i=16.30$,
and $z=16.47$ AB mag. Using the $T_{\rm eff}$, $\log$ g, and mass measurements from \citet{zuckerman07} and the model atmospheres
from \citet{koester05}, we estimate the theoretical colors for GD 362 and use these to constrain the solid angle $\pi (R/D)^2$,
which relates the flux at the surface of the star to that received at Earth. The mass radius relation of \citet{wood92}
gives a radius of 0.0109 $R_\odot$ for a 0.73 $M_\odot$ white dwarf with thin hydrogen layers (thin layers because GD 362 is
helium-rich). Using this radius, the distance for the model atmosphere fit parameters presented in \citet{zuckerman07} is
52.2 pc, entirely consistent with our parallax observations.

Instead, if we use our parallax measurement directly to constrain the radius of the star, we can use the surface gravity and the
radius to constrain the mass. Replacing the distance in the solid angle $\pi (R/D)^2$ with 50.6 pc, we find a radius of
7.357 $\times 10^8$ cm, or 0.0106 $R_\odot$ for GD 362. Using the surface gravity of $\log$ g = 8.24, we estimate a mass
of 0.71 $M_\odot$. Again this mass estimate is entirely consistent with the mass estimate of 0.73 $\pm$ 0.02
$M_\odot$ by \citet{zuckerman07}.

\section{Discussion}

GD 362 is only slightly more massive than the average mass for the field DA \citep[0.60 $\pm$ 0.13 $M_\odot$][]{liebert05} and
DB \citep[0.60 $\pm$ 0.07 $M_\odot$][]{voss07} white dwarfs. Therefore, there is no evidence of and no need to invoke a binary
merger scenario to explain this star and its surrounding debris disk. The photospheric abundances of metals in GD 362 are also
several orders of magnitude different from the nucleosynthetic predictions of a merger event \citep{zuckerman07}.
Dropping GD 362 from the list of massive stars, there are no known white dwarfs with $M>0.9M_\odot$ that host dust disks. However,
massive stars comprise only about 15\% of the local white dwarf population \citep{liebert05}, and only a small number of these
stars have been searched for excess infrared radiation from dust disks. A large survey of massive white dwarfs will be useful
to constrain the fraction with debris disks.

Using the initial-final mass relations by \citet{dobbie06, williams07, kalirai08}, we estimate the progenitor
of GD 362 to be a $3.0-3.3M_\odot$ star. The main sequence lifetime of a solar-metallicity $3.0 M_\odot$ star is 320 - 650 Myr,
depending on the assumptions on convecting overshooting (M. H. Pinsonneault 2008, priv. comm.). The white dwarf cooling age of GD 362
is $\approx$ 700-800 Myr, therefore the total main sequence plus white dwarf age of GD 362 is 1 - 1.5 Gyr. 
Using the proper motion and distance, we estimate that GD 362 has a tangential velocity of 51 km s$^{-1}$. Including the
radial velocity measurement from \citet[][after subtracting the gravitational redshift component]{zuckerman07}, we estimate
that GD 362 has U, V, and W velocities of 57.7, 0.5, and $-$5.3 km s$^{-1}$, respectively. These velocities are
consistent with thin disk membership. GD 362 seems just like all the other known white dwarfs with debris disks \citep{kilic08},
except that it has a mixed H/He atmosphere, and it has more metals than the rest of them.

\citet{jura07} modelled the mid-infrared spectral energy distribution of GD 362 with a flat disk between 12 and 50 stellar radii
and warped disk between 50 and 70 stellar radii. Using our radius estimate of 0.0106 $R_\odot$, we can convert these numbers into
observables. The dust disk lies between 0.13 and 0.74 $R_\odot$. In addition, based on a distance of 25 pc, the mass of circumstellar
dust was estimated to be 3 $\times 10^{17}$ g. Since the real distance to GD 362 is about twice as large, the actual dust mass is four
times higher, or $\approx 1.2 \times 10^{18}$ g. Of course, this is only a tiny fraction of the metals \citep[$\geq10^{22}$ g,][]{zuckerman07}
present in the convective envelope of the star.

\section{Conclusions}

We obtained trigonometric parallax observations of GD 362 over seven epochs separated by 1.3 years. Our direct distance measurement of
$50.6^{+3.5}_{-3.1}$ pc is incompatible with GD 362 being as massive as 1.2$M_\odot$. Using the parallax to constrain the radius
and the mass, we find that GD 362 has a mass $\approx$ 0.71 $M_\odot$. The model atmosphere analysis presented in \citet{zuckerman07}
demonstrated that GD 362 has significant amounts of helium in its photosphere and it has $M=0.73\pm 0.02 M_\odot$.
Our parallax observations confirm their result.

\acknowledgements
MK is grateful to the Ohio State University Astronomy Department, and especially to K. Stanek,
for large amounts of time allocated on the MDM 2.4m Telescope. Support for this work was provided by NASA
through the Spitzer Space Telescope Fellowship Program, under an award from Caltech. JRT acknowledges support
from the National Science Foundation through grants AST-0307413 and AST-0708810. We thank the MDM Observatory
staff for their excellent support, and J. Subasavage for a quick and constructive referee report.

\clearpage

\begin{deluxetable}{lrrrrr}
\tablewidth{0pt}
\tablecolumns{6}
\tablecaption{Journal of Observations}
\tablehead{
\colhead{Mean Date} &
\colhead{$N$} &
\colhead{First HA} &
\colhead{Last HA} &
\colhead{$p_X$} &
\colhead{$p_Y$} \\
}
\startdata
2007 May 27 & 27 &  $+$0:17  &  $+$2:01  & $0.31$ & $0.83$ \\
2007 Jul  4 & 15 &  $+$0:09  &  $+$1:04  & $-0.32$ & $0.84$ \\
2007 Sep 26 &  7 &  $+$1:36  &  $+$1:59  & $-0.99$ & $-0.12$ \\
2008 Mar 30 &  8 &  $-$0:06  &  $+$0:13  & $0.96$ & $0.21$ \\
2008 May 14 & 42 &  $+$0:17  &  $+$1:34  & $0.50$ & $0.75$ \\
2008 Jun 23 &  4 &  $+$0:36  &  $+$0:48  & $-0.15$ & $0.87$ \\
2008 Sep  6 & 14 &  $+$0:38  &  $+$1:29  & $-0.99$ & $0.17$ \\
\enddata
\tablecomments{Summary of the parallax observations.  Columns 1 and
2 give the mean date of the observing run and the number of parallax images
used in the analysis.  Columns 3 and 4 give the easternmost and westernmost
hour angles of the images used from that run, in hours and minutes.  
Columns 5 and 6 give the
mean parallax factors in $X$ (eastward) and $Y$ (northward).}
\end{deluxetable}

\clearpage
\begin{figure}
\hspace*{-0.5in}
\includegraphics[angle=-90,scale=.7]{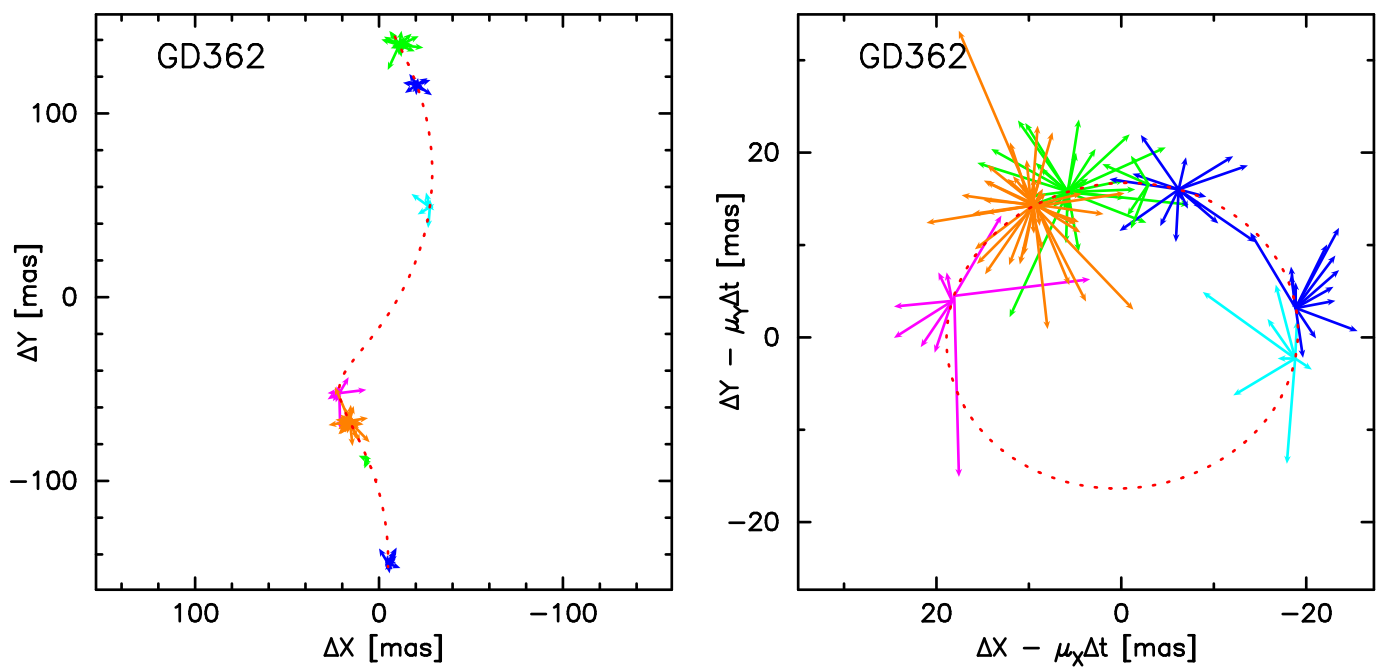}
\caption{Left panel: The trajectory of GD 362 on the
sky. Right panel: The same trajectory with the proper motion taken
out. The tip of each arrow is the position from a single image, and
the tail is the computed location based on the fitted trajectory including
zero point, proper motion, and parallax.}
\end{figure}

\end{document}